\documentclass[pdf]{iucr}              

                   \def\href#1{\relax}\let\foo\caption
\let\caption\foo

     \paperprodcode{a000000}      
     \paperref{xx9999}            
     \papertype{IU}               

     \paperlang{english}          
     \journalyr{2003}
     \journalreceived{\relax}
     \journalaccepted{\relax}
     \journalonline{\relax}
\usepackage{SIunits}
\usepackage{amsmath}
\begin{document}
\title{Coherence properties of focused X-ray beams at high brilliance synchrotron sources}
\author[a]{Andrej}{Singer}
\cauthor[a,b]{Ivan}{Vartanyants}{Ivan.Vartaniants@desy.de}{}
\aff[a]{Deutsches Elektronen-Synchrotron DESY, Notkestra\ss{}e 85, D-22607 Hamburg, Germany}
\aff[b]{National Research Nuclear University ''MEPhI'', 115409 Moscow, Russia}
\keyword{X-Ray Lenses, Partial Coherence}
\maketitle


\begin{abstract}
An analytical approach describing properties of focused partially coherent X-ray beams is presented.
The method is based on the results of statistical optics and gives both the beam size and transverse coherence length at any distance behind an optical element.
In particular, here we consider Gaussian Schell-model beams and thin optical elements.
Limiting cases of incoherent and fully coherent illumination of the focusing element are discussed.
The effect of the beam defining aperture, typically used in combination with focusing elements at synchrotron sources to improve transverse coherence, is also analyzed in detail.
As an example the coherence properties in the focal region of compound refractive lenses at the PETRA III synchrotron source are analyzed.
\end{abstract}

\section{Introduction}

While ultimate storage rings, being diffraction limited X-ray sources, are still under development \cite{BeiNIMA2010}, present third generation synchrotrons are partially coherent sources \cite{VartanyantsNJP2010}.
The construction of these sources initiated developments of new research areas, which utilize partial coherence of the X-ray radiation.
Most prominent among these techniques are coherent X-ray diffractive imaging (CXDI) \cite{Vartanyants2010,CN2010,MancusoJB2010,VartanyantsBook} and X-ray photon correlation spectroscopy (XPCS) \cite{GruebelJAC2004}.
In CXDI static real space images of the sample are obtained by phase retrieval techniques \cite{F1982}, whereas in XPCS dynamics of a system are explored by correlation techniques \cite{G2007}.

The key feature of all coherence based methods is the interference of the field scattered by different parts of the sample.
As such, spatial coherence across the sample is essential and understanding the coherence properties of the incoming X-Ray beams generated at new generation synchrotron sources is of vital importance for the scientific community.
A detailed knowledge of the coherence properties can even be used to improve the resolution obtained in the CXDI phase retrieval \cite{WWQ2009}.

\begin{figure}
  \scalebox{0.13}{\includegraphics{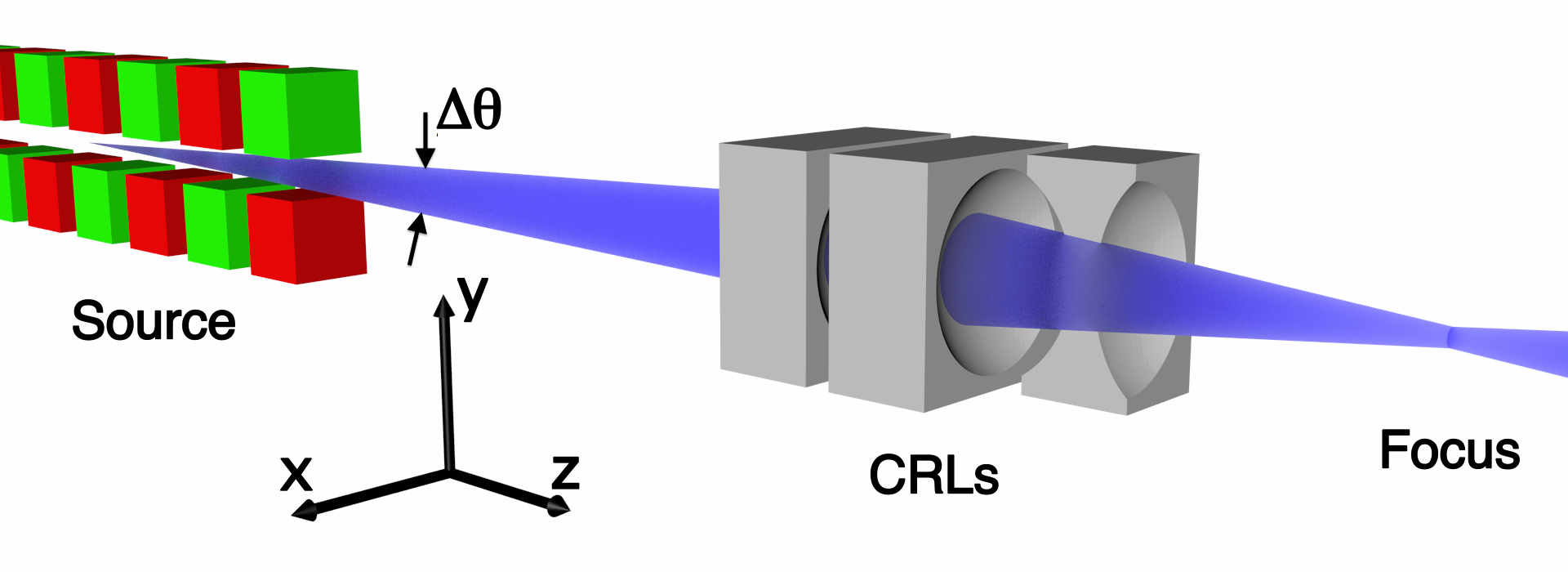}}
  \caption{Partially coherent radiation is generated in the undulator and is focused by a stack of CRLs.
Intensity and coherence properties of the focused radiation are considered.
The last lens is cut to indicate the structure of a single CRL.
  }
  \label{fig:Propagation_Lenses}
\end{figure}

For scientific applications at the nanoscale, beam sizes from tens to hundreds of nanometers with high flux densities are required.
These can be achieved by an effective use of focusing elements.
Nowadays several techniques to focus X-ray beams at third and fourth generation sources are used, such as Kirkpatrick-Baez (KB) mirrors \cite{MHS2010}, Fresnel zone plates \cite{SA2010}, bent crystals in Bragg geometry \cite{ZhuAPL2012}, and compound refractive lenses (CRL) \cite{SKS1996,SKH2003}.
A typical focusing scheme is shown in Figure \ref{fig:Propagation_Lenses}.
Synchrotron radiation is generated in the undulator and a focusing element consisting of a stack of CRLs focuses the beam.
In this paper we describe the propagation of partially coherent X-ray radiation through such a focusing system and determine its size and coherence properties at any position downstream from the CRL.
Our results can be naturally generalized to other type of focusing elements such as Fresnel zone plates.

Synchrotron sources are generally considered as incoherent sources, since different electrons in the electron bunch radiate independently in the frame moving with the electrons.
Due to relativistic effects, in the laboratory frame the radiation is confined to a narrow cone of angles $\Delta\theta\le1/2\gamma$ (see Figure \ref{fig:Propagation_Lenses}), where $\gamma$ is the Lorentz factor.
This relativistic confinement implies an effective degree of transverse coherence at the source, as totally incoherent sources radiate into all directions \cite{G1985}.

The transverse coherence area $\Delta x\Delta y$ of a synchrotron source can be estimated from Heisenberg's uncertainty principle \cite{MW1995} $\Delta x \Delta y \ge {\hbar^2}/{4\Delta p_x \Delta p_y}$, where $\Delta x,~\Delta y$ and $\Delta p_x,~\Delta p_y$ are the uncertainties in the position and momentum in the horizontal and vertical direction, respectively.
Due to the de Broglie relation $p=\hbar k$, where $k=2\pi/\lambda$, the uncertainty in the momentum $\Delta p$ can be associated with the source divergence $\Delta\theta$, $\Delta p=\hbar k\Delta\theta$, and the coherence area in the source plane is given by
\begin{equation}
\Delta x\Delta y\ge
 \left(\frac{\lambda}{4\pi}\right)^2\frac{1}{\Delta\theta_x\Delta\theta_y} 
 \label{eq:coherence_synchrotron}.
\end{equation}
Substituting typical values of the source divergence at a third generation synchrotron source \cite{petra3} in equation (\ref{eq:coherence_synchrotron}) we find the minimum transverse coherence length at the source to be about few micrometers.
With the source sizes of tens to hundreds microns \cite{petra3} it is clear that present third generation X-ray sources have to be described as partially coherent sources.

A useful model to describe the radiation properties of partially coherent sources is the Gaussian Schell-model (GSM) \cite{MW1995}.
This model has been applied for the analysis of the radiation field generated by optical lasers \cite{G1980}, third generation synchrotron sources (see for example \cite{HK1994,VartanyantsNJP2010} and references therein) and X-ray free-electron lasers \cite{SVK2008,RSW2011,VSM2011,SSG2011}.
The problem of propagation of partially coherent radiation through thin optical elements (OE)  in the frame of GSM was widely discussed in optics \cite{TF1986,YH1987}.
However, the propagation of partially coherent radiation through the focusing elements with finite apertures was not considered before.
For X-ray beamlines at third generation synchrotron sources such focusing elements are especially important.
In this work we propose a general approach to describe propagation of partially coherent radiation through these beamlines.

The paper is organized as follows.
We start with a short introduction to the optical coherence theory with the special focus on third generation synchrotron radiation sources in section 2.
Section 3 describes the propagation of partially coherent X-ray radiation through thin focusing elements.
Diffraction limited focus and infinite apertures are described in sections 4 and 5.
In section 6 coherence properties of the focused X-ray beams at PETRA III are analyzed.
The paper is concluded with a summary and outlook.

\section{Coherence: Basic Equations}

\subsection{Correlation functions and propagation in free-space}
The theory of partially coherent fields is based on the treatment of correlation functions of the complex wave field \cite{MW1995}.
The concept of optical coherence is often associated with interference phenomena, where the Mutual Coherence Function (MCF)
\footnote{
Here we restrict ourselves to stationary radiation fields, which are generated at third generation synchrotron sources.
}
\begin{equation}
  \Gamma(\textbf r_1,\textbf r_2;\tau)
  =\left\langle E^*(\textbf r_1,t)E(\textbf r_2,t+\tau)\right\rangle,
  \label{eq:MCF}
\end{equation}
plays the main role.
It describes the correlations between two complex values of the electric field $E^*(\textbf r_1,t)$ and $E(\textbf r_2,t+\tau)$ at different points $\textbf r_1$ and $\textbf r_2$ and different times $t$ and $t+\tau$.
The brackets $\langle\cdots\rangle$ denote the time average.

When we consider propagation of the correlation function of the field in free space, it is convenient to introduce the cross-spectral density function (CSD), $W(\textbf r_1,\textbf r_2;\omega)$, which is defined as the Fourier transform of the MCF \cite{MW1995}
\begin{equation}
  W(\textbf r_1,\textbf r_2;\omega) =\int\Gamma(\textbf r_1,\textbf r_2;\tau)e^{-i\omega\tau}\mbox d\tau,
  \label{eq:WCT1}
\end{equation}
where $\omega$ is the angular frequency of the radiation.
By definition, when the two points $\textbf r_1$ and $\textbf r_2$ coincide, the CSD represents the spectral density of the radiation field,
\begin{equation}
  S(\textbf r;\omega) = W(\textbf r,\textbf r;\omega).
  \label{eq:SD}
\end{equation}
The normalized CSD is known as the spectral degree of coherence (SDC)
\begin{equation}
  \mu(\textbf r_1,\textbf r_2;\omega) =  \frac{W(\textbf r_1,\textbf r_2;\omega)}{\sqrt{S(\textbf r_1;\omega)S(\textbf r_2;\omega)}}.
  \label{eq:SDC}
\end{equation}
For all values of  $\textbf r_1,\textbf r_2$ and $\omega$ the SDC satisfies $|\mu (\textbf r_1,\textbf r_2;\omega)|\leq 1$.
The modulus of the SDC can be measured in interference experiments as the contrast of the interference fringes \cite{SVK2008,VSM2011,SSG2011}.

To characterize the transverse coherence properties of a wave field by a single number, the global degree of transverse coherence can be introduced as \cite{SSY2008,VartanyantsNJP2010}
\begin{eqnarray}
  \zeta(\omega) = \frac{\int \left|W(\textbf r_1,\textbf r_2;\omega)\right|^2\mbox d\textbf r_1\mbox d\textbf r_2 }
  {\left(\int S(\textbf r;\omega)\mbox d\textbf r\right)^2 }.
\label{eq:NDC}
\end{eqnarray}%
According to its definition the values of the parameter $\zeta (\omega )$ lie in the range 0$\leq \zeta (\omega )\leq 1$, where $\zeta(\omega)=1$ and $\zeta(\omega)=0$ characterizes fully coherent and incoherent radiation, respectively.

In  the following we will apply the concept of correlation functions to planar secondary sources \cite{MW1995}, where the CSD of the radiation field is given in the source plane at $z_0=0$ with the transverse coordinates $\textbf s$, $W(\textbf s_1,\textbf s_2,z_0;\omega)$.
The propagation of the CSD from the source plane at $z_0$ to the plane at a distance $z$ from the source is governed by the following expression \cite{MW1995}

\begin{equation}
  \begin{split}
  W(&\textbf u_1,\textbf u_2,z;\omega)=\\
  &\int W(\textbf s_1,\textbf s_2,z_0;\omega)
  P_z^{\ast }(\textbf u_1,\textbf s_1;\omega)
  P_z(\textbf u_2,\textbf s_2;\omega)
  \mbox d\textbf s_1\mbox d\textbf s_2,
 \end{split}
 \label{eq:CSD_propagation}
\end{equation}
where $W(\textbf u_1,\textbf u_2,z;\omega )$ is the propagated CSD in the plane $z$, and $P_z(\textbf u,\textbf s;\omega )$ is the propagator.
The integration is performed in the source plane.
For partially coherent X-ray radiation at third generation synchrotron sources it is typically sufficient to use the Fresnel propagator \cite{G2005}, which is given by
\begin{equation}
  P_z(\textbf u,\textbf s;\omega)=\frac{ke^{ikz}}{2\pi iz}
  \exp\left(ik\frac{|\textbf u-\textbf s|^2}{2z}  \right).
  \label{eq:Fresnel_propagator}
\end{equation}

\subsection{Gaussian Schell-model sources}

The CSD of a GSM source positioned in the plane at $z_0$ is expressed as \cite{MW1995}\footnote{In this equation and below we omit the frequency dependence $\omega$ for brevity.
The GSM will be applied to narrow-bandwidth radiation, where $\omega$ is the average frequency.
}
\begin{eqnarray}
  W(\textbf s_1,\textbf s_2;z_0)=\sqrt{S(\textbf s_1)}\sqrt{ S(\textbf s_2)}\mu(\textbf s_2-\textbf s_1),
\label{eq:GSMa}
\end{eqnarray}
where the spectral density and the SDC in the source plane are Gaussian functions
\begin{equation}
  \begin{split}
  S(\textbf s) &=S_{0}
  \exp\left(-\frac{s_x^2}{2\sigma_x^2}-\frac{s_y^2}{2\sigma_y^2}\right) \\
  \mu(\textbf s_2-\textbf s_1) &=
  \exp\left(-\frac{(s_{2x}-s_{1x})^2}{2\xi_x^2}-\frac{(s_{2y}-s_{1y})^2}{2\xi _y^2}\right).
  \end{split}
  \label{eq:GSMb}
\end{equation}
Here $S_0$ is a normalization constant, and the parameters $\sigma _{x,y}$ and $\xi _{x,y}$ define the source size and transverse coherence length in the source plane in $x$- and $y$- directions, respectively.
Below all values are presented as root mean square (rms) values, if not stated differently.

The expression of the CSD function in the form of equation (\ref{eq:GSMa}), is based on the definition of the SDC (\ref{eq:SDC}).
In the GSM the main approximations are that the source is spatially uniform, i.e. $\mu(\textbf{s}_1,\textbf{s}_2)= \mu(\textbf s_2-\textbf s_1)$, and all functional dependencies are described by Gaussian functions.

%

The CSD $W(u_1,u_2;z)$ at the distance $z$ from the source can be calculated using integration of equation \eqref{eq:CSD_propagation} with the Fresnel propagator \eqref{eq:Fresnel_propagator} \cite{MW1995}%
\footnote{
It is noteworthy that the CSD of a GSM source can be factorized into two transverse components.
We will present calculations for one transverse direction and will drop the subscript for brevity.
}
\begin{eqnarray}
  \begin{split}
  W(u_1,u_2&,z)=\\
  &\frac{\sqrt{S_0}\textrm{e}^{i\psi_{12}(z)}}{\Delta(z)}
\exp \left( -\frac{u_1^2+u_2^2}{4\Sigma^2(z)}
-\frac{(u_2-u_1)^2}{2\Xi^2(z)}\right),
\end{split}
\label{eq:GSM_pfs}
\end{eqnarray}
where
\begin{equation}
	\Sigma(z)=\sigma\Delta(z),\quad\textrm{and} \quad\Xi(z)=\xi\Delta(z)
\label{eq:SigXi}
\end{equation}
are the beam size and transverse coherence length at the distance $z$ from the source.
The parameter
\begin{eqnarray}
  \Delta(z)=\sqrt{1+\left( \frac{z}{z_\textrm{eff}}\right) ^{2}}
\label{eq:Delta}
\end{eqnarray}
is the expansion coefficient and
\begin{eqnarray}
\psi_{12}(z) = k\frac{u_2^2-u_1^2}{2R(z)},\quad R(z)=z\left[1+\left( \frac{z_\textrm{eff}}{z}\right) ^{2}\right]
\label{eq:R}
\end{eqnarray}
are the phase and radius of curvature of the GSM beam.
In equations (\ref{eq:Delta}, \ref{eq:R}) the effective distance
\begin{equation}
z_\textrm{eff} = 2k\sigma^2\zeta
\label{eq:zeff}
\end{equation}
has been introduced \cite{GburOptComm2001,VartanyantsNJP2010}.
At that distance the expansion coefficient is equal to $\Delta(z_\textrm{eff})=\sqrt{2}$.
In the limit of a spatially coherent source, $\zeta=1$, the effective distance $z_\textrm{eff}$ coincides with the Rayleigh length $z_R=2k\sigma^2$, which is often introduced in the theory of optical Gaussian beams \cite{ST1991}.
It is noteworthy, that the CSD of the beam downstream of the source is not homogeneous, i.e. $\mu(u_1,u_2)\neq\mu(u_2-u_1)$ due to the phase factor $\psi_{12}(z)$ in equation \eqref{eq:GSM_pfs}.

It is important to note here that in the frame of the GSM the coherence properties of the beam at any position along the beamline containing OEs (see Figure \ref{fig:Propagation_Lenses}) will be described by the same equation \eqref{eq:GSM_pfs} with different meaning of the parameters $\Sigma(z)$, $\Xi(z)$, $\psi_{12}(z)$, and $\Delta(z)$.

The global degree of coherence of a GSM source can be expressed as \cite{VartanyantsNJP2010}
\begin{equation}
\zeta = \frac{1}{\sqrt{1 + (2\Sigma(z)/\Xi(z))^2}}.
\label{eq:zeta_GSM}
\end{equation}
One important property of the GSM beams is that in the case of free space propagation the global degree of coherence remains constant (see equations (\ref{eq:SigXi},~\ref{eq:zeta_GSM})).

\subsection{Propagation through optical elements}

The propagation of the CSD through a thin OE can be described by a complex valued transmission function
$T(\textbf u)$ \cite{G1985}
\begin{equation}
  \tilde W(\textbf u_1,\textbf u_2,z)=W(\textbf u_1,\textbf u_2,z)T^*(\textbf u_1)T(\textbf u_2),
  \label{eq:CSD_thin_lens}
\end{equation}
where $W(\textbf u_1,\textbf u_2,z)$ and $\tilde W(\textbf u_1,\textbf u_2,z)$ are the CSDs incident on and just behind the OE.
It is interesting to note that a thin OE described by a transmission function $T(\textbf u)$ does not change the transverse coherence properties in its plane.
It can be readily seen from equations (\ref{eq:SDC},~\ref{eq:CSD_thin_lens}) that the modulus of the SDC in front $|\mu(\textbf u_1,\textbf u_2,z)|$ and behind $|\tilde\mu(\textbf u_1,\textbf u_2,z)|$ the lens are the same $\vert\tilde\mu(\textbf u_1,\textbf u_2,z)\vert = \vert \mu(\textbf u_1,\textbf u_2,z)\vert$.
This also implies that according to equation \eqref{eq:GSM_pfs} the coherence length $\Xi(z)$ will be preserved.

In general, the propagation of partially coherent radiation through a beamline with a thin OE can be
performed in the following steps.
First, the CSD $W(\textbf s_1,\textbf s_2,z_0)$ at the source is defined.
The propagation of the CSD $W(\textbf s_1,\textbf s_2,z_0)$ from the source to the first OE positioned at $z_\textrm{L}$ can be described by equation (\ref{eq:CSD_propagation}).
For the propagation of the CSD $W(\textbf u_1,\textbf u_2,z_\textrm{L})$ through the OE equation (\ref{eq:CSD_thin_lens}) can be utilized.
Finally, the coherence properties at any position $z_1$ downstream of this OE are obtained using equation (\ref{eq:CSD_propagation}).
The extension of this procedure to simulate the propagation of partially coherent radiation through a beamline containing several OEs is straightforward, provided each OE can be well described in the frame of the thin OE approximation.
Below we will implement this scheme for a simple beamline geometry containing an undulator source described by a plane GSM source and a focusing element positioned at a distance $z_\textrm{L}$ downstream of the source (see Figure \ref{fig:Propagation_Lenses}).

\section{Focusing of partially coherent x-ray beams}
\bigskip

\subsection{Compound Refractive Lenses}

As a focusing element we will consider a parabolic CRL \cite{LengelerAPL1999}.
The complex valued transmission function $T(\textbf u)$ of such a lens can be written in the form%
\footnote{In principle, the transmission function can depend on the frequency of radiation $\omega$.
However, for CRLs in the x-ray range for a bandwidth lower than $10^{-3}$ the frequency dependence can be neglected \cite{K2012}.}

\begin{equation}
T(\textbf u)=B(\textbf u)\exp \left( -i\frac{k\vert\textbf u\vert^{2}}{2f}\right).
\label{eq:transmission_thin_lensa}
\end{equation}
The function $B(\textbf u)$ defines the absorption and opening aperture of the lens and $f$ is its focal length \cite{ST1991,G2005}
\begin{equation}
f=\frac{R}{2\delta}.
\label{eq:f0}
\end{equation}
Here $\delta$ is the real part of the complex index of refraction \cite{BW1999} $n=1-\delta+i\beta$ that is on the order of $10^{-6}$ for X-Ray energies.
The parameter $\beta$ is the imaginary part of the refractive index and describes absorption.
Since $\delta$ is extremely small for x-rays, typically several lenses are stacked together (see Figure \ref{fig:Propagation_Lenses}) to reduce the focal length and improve the focusing properties of the lenses.
For a combination of $N$ lenses the focal length is given by
\begin{eqnarray}
f&=&\frac{1}{2\delta}\left(\sum_{i=1}^N\frac{1}{R_i}\right)^{-1},\label{eq:f}
\end{eqnarray}
where $R_i$ is the radius of $i$-th lens.
The above expression holds if the total arrangement of lenses fulfils the thin lens approximation.

Lens imperfections or aberrations, if present, can be taken into account by introducing additional phase factors in $B(\textbf u)$.
Here, we restrict ourselves to aberration free optics and assume that for a thin parabolic lens the opening aperture function can be described by a Gaussian function
\begin{equation}
B(\textbf u)=B_0\exp \left( -\frac{\vert\textbf u\vert^{2}}{4\Omega_0^{2}}\right),
\label{eq:transmission_thin_lensb}
\end{equation}
where $\Omega_0$ is the effective opening aperture due to absorption in the material of the lens defined through 
\begin{equation}
\Omega_0^2=\frac{f\delta}{2k\beta}.
\label{eq:Omega}
\end{equation}
The parameter $B_0$ describes the transmission of the lens in its center and satisfies $0<B_0\le1$.

It is important to note that often the opening aperture of the OE is determined not by the natural absorption but rather by the size of the lens or beam defining aperture in front of the lens.
When this additional aperture $\Omega_A$ is comparable with or smaller than the effective aperture $\Omega_0$ we introduce the total aperture
\begin{equation}
\frac1{\Omega^2}= \frac1{\Omega_0^2}+\frac{1}{\Omega_A^2}.
\label{eq:Omegatot}
\end{equation}
To simplify the analysis we consider here a Gaussian form of the additional aperture.
We show in the Appendix \ref{chap:A1} that the coherence properties of the focused radiation do not significantly change if a rectangular aperture of the corresponding size is used.

\subsection{Propagation of Gaussian Schell-model beams through focusing elements}

\begin{figure}
\centering
\scalebox{0.7}{\includegraphics{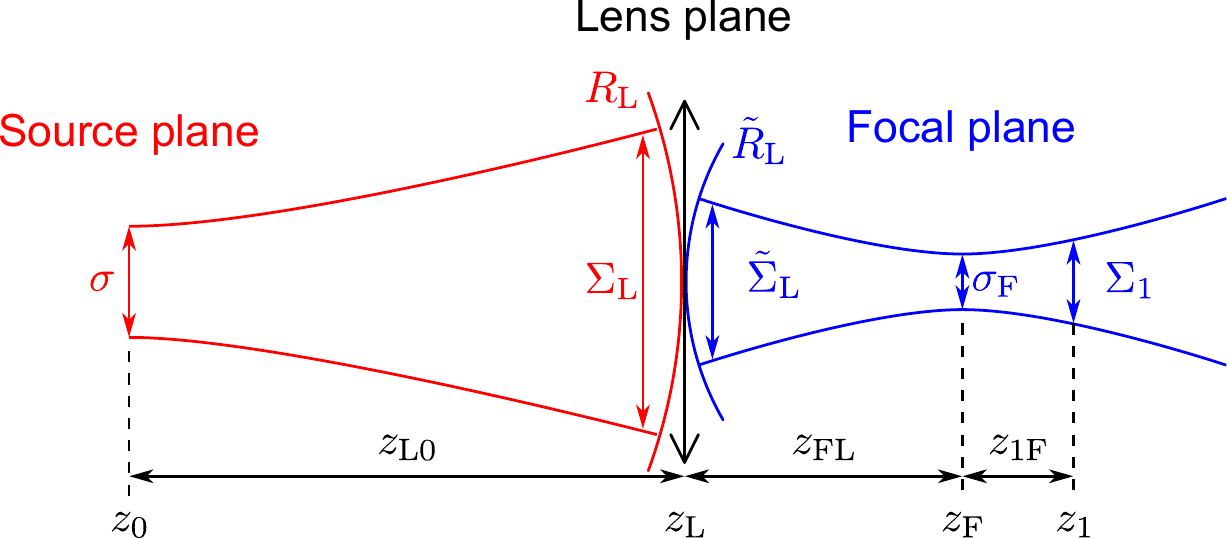} }
\caption{The propagation geometry. The source with a source size $\sigma$ is positioned at $z_0$.
Partially coherent radiation with the beam size $\Sigma_\textrm{L}$, coherence length $\Xi_\textrm{L}$, and radius of curvature $R_\textrm{L}$ is incident on the lens at $z_\textrm{L}$.
The beam size $\tilde\Sigma_\textrm{L}$ and the radius of curvature $\tilde R_\textrm{L}$ are modified by the transmission through the lens.
The beam is focused at the focal position $z_\textrm{F}$ with the focus size $\sigma_\textrm{F}$.
The beam parameters at an arbitrary position $z_1$ downstream of the lens are determined.
}
\label{fig:propagation_geometry}
\end{figure}

To simulate propagation of partially coherent radiation through a focusing element we will use the procedure outlined above (see Figure \ref{fig:propagation_geometry}).
The source at $z_0$ will be described in the frame of the Gaussian Schell-model with the CSD $W( s_1,s_2,z_0)$ defined in equations (\ref{eq:GSMa},\ref{eq:GSMb}).
The CSD function incident on the lens at $z_\textrm{L}$ is given by equation (\ref{eq:GSM_pfs}).
The parameters $\Sigma_\textrm{L}=\Sigma(z_\textrm{L})$, $\Xi_\textrm{L}=\Xi(z_\textrm{L})$, $\Delta_\textrm{L}=\Delta(z_\textrm{L})$, and  $R_\textrm{L}=R(z_\textrm{L})$ are the beam size, transverse coherence length, expansion coefficient, and radius of curvature incident on the lens at $z_\textrm{L}$, respectively (see equations (\ref{eq:SigXi},~\ref{eq:Delta},~\ref{eq:R})).
Substituting the lens transmission function introduced in equations (\ref{eq:transmission_thin_lensa},~\ref{eq:transmission_thin_lensb}) in equation (\ref{eq:CSD_thin_lens}) we can determine the CSD immediately behind the lens.
The beam behind a Gaussian lens with an opening aperture $\Omega$ can be again discribed by the GSM using equation \eqref{eq:GSM_pfs} with the modified beam size
\begin{equation}
    \frac{1}{\tilde\Sigma_\textrm{L}^2}=\frac1{\Sigma_\textrm{L}^2}+\frac1{\Omega^2},
\label{eq:SigmaBehind}
\end{equation}
radius of curvature
\begin{equation}
\frac1{\tilde R_\textrm{L}}=\frac1{R_\textrm{L}}-\frac1f,
    \label{eq:RBehind}
\end{equation}
and normalization constant $\sqrt{S_0}B_0/\Delta_\textrm{L}$ (see Figure \ref{fig:propagation_geometry}).
As mentioned earlier, in the thin lens approximation the coherence length $\Xi_\textrm{L}=\Xi(z_\textrm{L})$ is not modified while the transmission of the incident beam through the lens.

If the beam size $\tilde \Sigma_\textrm{L}$ is reduced due to a finite aperture $\Omega$ of the lens the global degree of transverse coherence \eqref{eq:zeta_GSM} behind the lens can be defined as
\begin{equation}
\zeta_\textrm{F}=\frac{1}{\sqrt{1 + \left( 2 \tilde\Sigma_\textrm{L}/\Xi_\textrm{L}\right)^2}}.
\label{eq:zeta_F}
\end{equation}
For partially coherent Gaussian beams this value will be constant at all positions downstream of the lens.

\begin{figure}
\centering
\scalebox{0.4}{\includegraphics{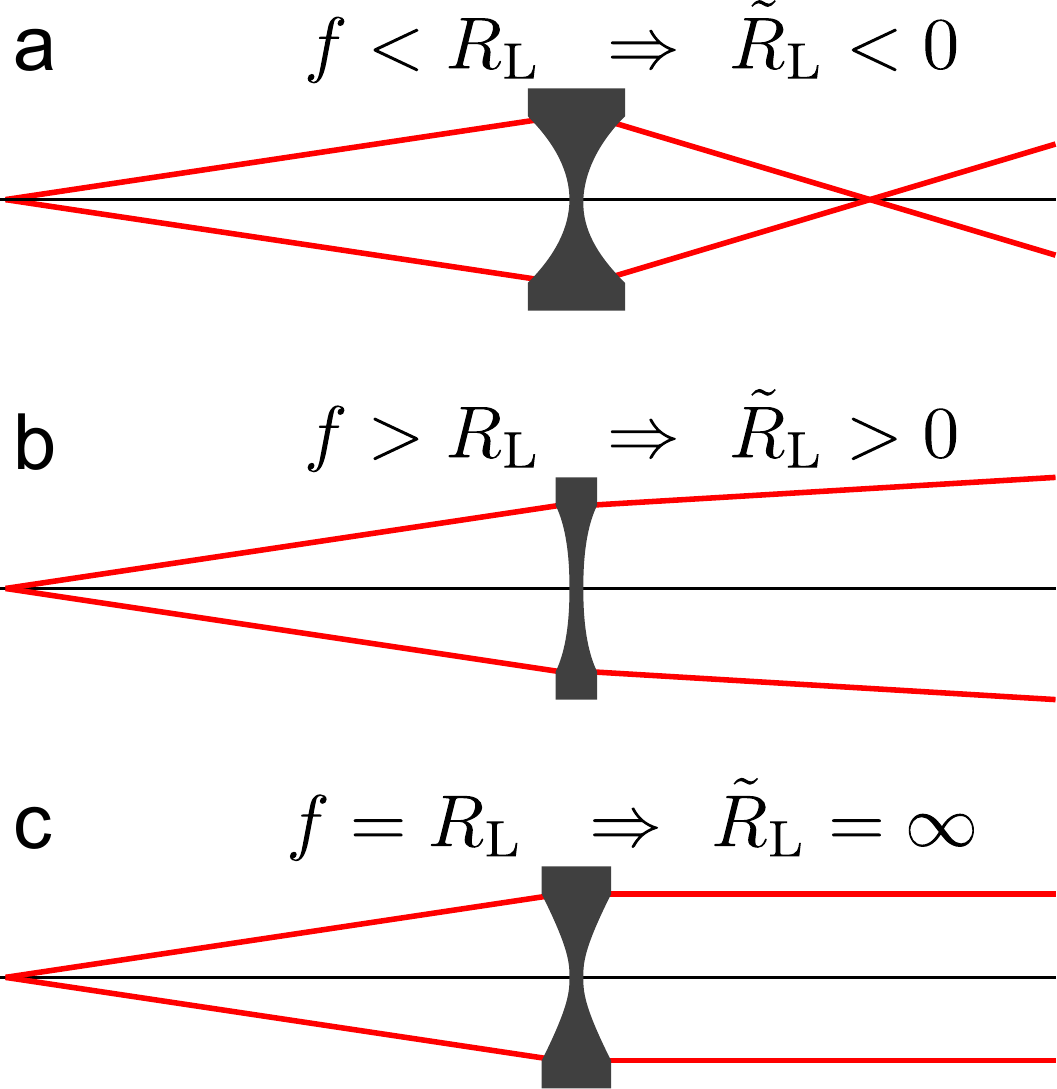} }
\caption{Different focusing geometries. (a) The beam is focused. (b) The divergence of the beam is reduced, but the beam is not focused.
(c) The beam is collimated.}
\label{fig:SketchFocus}
\end{figure}

It is well known, that the focusing properties of a lens are determined by the focal length $f$.
Depending on the sign of $f$ the lens acts as a focusing $f>0$ or a defocusing $f<0$ optical element.
We consider a lens with the focal length $f>0$, which according to equation \eqref{eq:RBehind} reduces the radius of curvature of the incident beam $\tilde R_\textrm{L}$.
If the focal length is smaller than the curvature of the incident beam, $f<R_\textrm{L}$, then according to equation \eqref{eq:RBehind} the radius of curvature behind the lens $\tilde R_\textrm{L}$ is negative and the beam is focused downstream of the lens (see Figure \ref{fig:SketchFocus} (a)).
In the opposite case of $f>R_\textrm{L}$ equation \eqref{eq:RBehind} yields a positive radius of curvature $\tilde R_\textrm{L}$ behind the lens.
The divergence of the beam is reduced, however, the beam is not focused and a virtual focus lies upstream from the lens (see Figure \ref{fig:SketchFocus} (b)).
If $f=R_\textrm{L}$ the radius of curvature behind the lens is infinite, which means that the beam is collimated (see Figure \ref{fig:SketchFocus} (c)).

\subsection{Coherence properties of the beam behind the focusing element}

To determine the beam properties in the focal plane one can apply the general propagation formula (\ref{eq:CSD_propagation}) to the radiation immediately behind the lens.
However, it is more convenient to use the optics reciprocity theorem \cite{BW1999}.
We assume that a source is located at the focal position $z_\textrm{F}$ and the beam is characterized by its CSD in the frame of the GSM by equations (\ref{eq:GSMa},~\ref{eq:GSMb}), with its source size and coherence length in the focus given by the parameters $\sigma_\textrm{F}$ and $\xi_\textrm{F}$, respectively.
To calculate these parameters we propagate partially coherent beam from the focal position $z_\textrm{F}$ backwards to the lens position $z_\textrm{L}$ using equation \eqref{eq:GSM_pfs} and compare it with the CSD function corresponding to the radiation transmitted through the lens.
The parameters of the focus satisfying this boundary condition are given by (see Appendix \ref{chap:A2} for details)
\begin{equation}
  \sigma_\textrm{F}=\frac{\tilde\Sigma_\textrm{L}}{ \sqrt{1+(Z_\textrm{L}/\tilde R_\textrm{L})^2}}
  \label{eq:sigma_f}
\end{equation}
\begin{equation}
  \xi_\textrm{F}=\frac{\Xi_\textrm{L}}{\sqrt{1+(Z_\textrm{L}/\tilde R_\textrm{L})^2}},
  \label{eq:xi_f}
\end{equation}
where we introduced  $Z_\textrm{L}=2k\tilde\Sigma_\textrm{L}^2\zeta_\textrm{F}$, which is similar to the effective distance $z_\textrm{eff}$ defined in equation \eqref{eq:zeff}.

The distance $z_\textrm{FL}$ from the lens to the focus is given by (see Appendix \ref{chap:A2} for details)
\begin{equation}
    z_\textrm{FL}=-\frac{\tilde R_\textrm{L}}{1+\left({\tilde R_\textrm{L}}/{Z_\textrm{L}}\right)^2}.
\label{eq:FocalPosition}
\end{equation}
In this model the radius of curvature of the radiation in the focus is infinitely large and the phase $\psi_{12}(z)$ term in equation \eqref{eq:GSM_pfs} vanishes.
It is readily seen from equations (\ref{eq:RBehind},~\ref{eq:FocalPosition}) that the focal position coincides with the focal length of the lens $z_\textrm{FL}=f$ only if the radius of curvature $R_\textrm{L}$ incident on the lens is much larger than the focal length $R_\textrm{L}\gg f$ and $Z_\textrm{L}\gg f$, that is typically the case for the third generation x-ray synchrotron sources.

The depth of focus $\Delta f$ is the region along the optical axis, where the beam size is smaller than the focus size multiplied by $\sqrt{2}$.
It is typically defined through the Rayleigh length for coherent Gaussian beams \cite{ST1991} and can be extended to partially coherent beams introducing the effective distance $z_\textrm{eff}^\textrm{F}=2k\sigma_\textrm{F}^2\zeta_\textrm{F}$ in the focus
\begin{equation}
	\Delta f = 2 z_\textrm{eff}^\textrm{F}.
\end{equation}

After the position of the focus and transverse coherence properties in the focus have been obtained, it is possible to calculate the CSD at any position  $z_1$ downstream of the lens applying equations (\ref{eq:GSM_pfs}-\ref{eq:zeff}).
In these equations the source size $\sigma$, transverse coherence length at the source $\xi$, and the global degree of coherence $\zeta$ are replaced by the values $\sigma_\textrm{F}$, $\xi_\textrm{F}$, and $\zeta_\textrm{F}$ in the focus from equations (\ref{eq:zeta_F},~\ref{eq:sigma_f},~\ref{eq:xi_f}).
The distance $z$ from the source to the observation plane is replaced by $z_{1\textrm{F}}=z_1-z_\textrm{F}$, which is the distance between the observation plane at $z_1$ and the focus at $z_\textrm{F}$ (see Figure \ref{fig:propagation_geometry}).
Below the limits of a fully coherent or diffraction limited focus as well as rather incoherent focus will be discussed.

\section{Diffraction limited focus}

We will consider now a strongly focusing lens, which substantially increases the flux density in the focus and is especially interesting for practical applications.
According to equation \eqref{eq:sigma_f} a small focal size $\sigma_\textrm{F}$ occurs, when the denominator in equation \eqref{eq:sigma_f} is large.
This is equivalent to the condition that the beam curvature behind the lens $\tilde R_\textrm{L}\ll Z_\textrm{L}$.
In this limit we obtain from equation \eqref{eq:sigma_f}
\begin{equation}
	\sigma_\textrm{F} = \frac{z_\textrm{FL}}{2k\tilde\Sigma_\textrm{L}\zeta_\textrm{F}}.
\label{eq:sigmaf1}
\end{equation}
Here we also used the fact, that in the same limit of $\tilde R_\textrm{L}\ll Z_\textrm{L}$ according to equation \eqref{eq:FocalPosition} the focal distance $z_\textrm{FL}\to -\tilde R_\textrm{L}$ and can be expressed as
\begin{equation}
z_\textrm{FL}=\frac{fR_\textrm{L}}{R_\textrm{L}-f}.
\end{equation}
Introducing the diffraction limited focus size $\sigma_\textrm{dl}=z_\textrm{FL}/(2k\Omega)$ equation \eqref{eq:sigmaf1} can be presented as
\begin{equation}
	\sigma_\textrm{F} = \sigma_\textrm{dl} \left( \frac{\Omega} {\tilde\Sigma_\textrm{L}}\right)\left( \frac{1} {\zeta_\textrm{F}}\right).
\label{eq:DiffLim}
\end{equation}
The diffraction limit can be equivalently written as $\sigma_\textrm{dl}=\lambda/(4\pi \textrm{NA})$, where NA=$\Omega/z_\textrm{FL}$ is the numerical aperture of the lens.

\begin{figure}
\centering
\scalebox{0.43}{\includegraphics{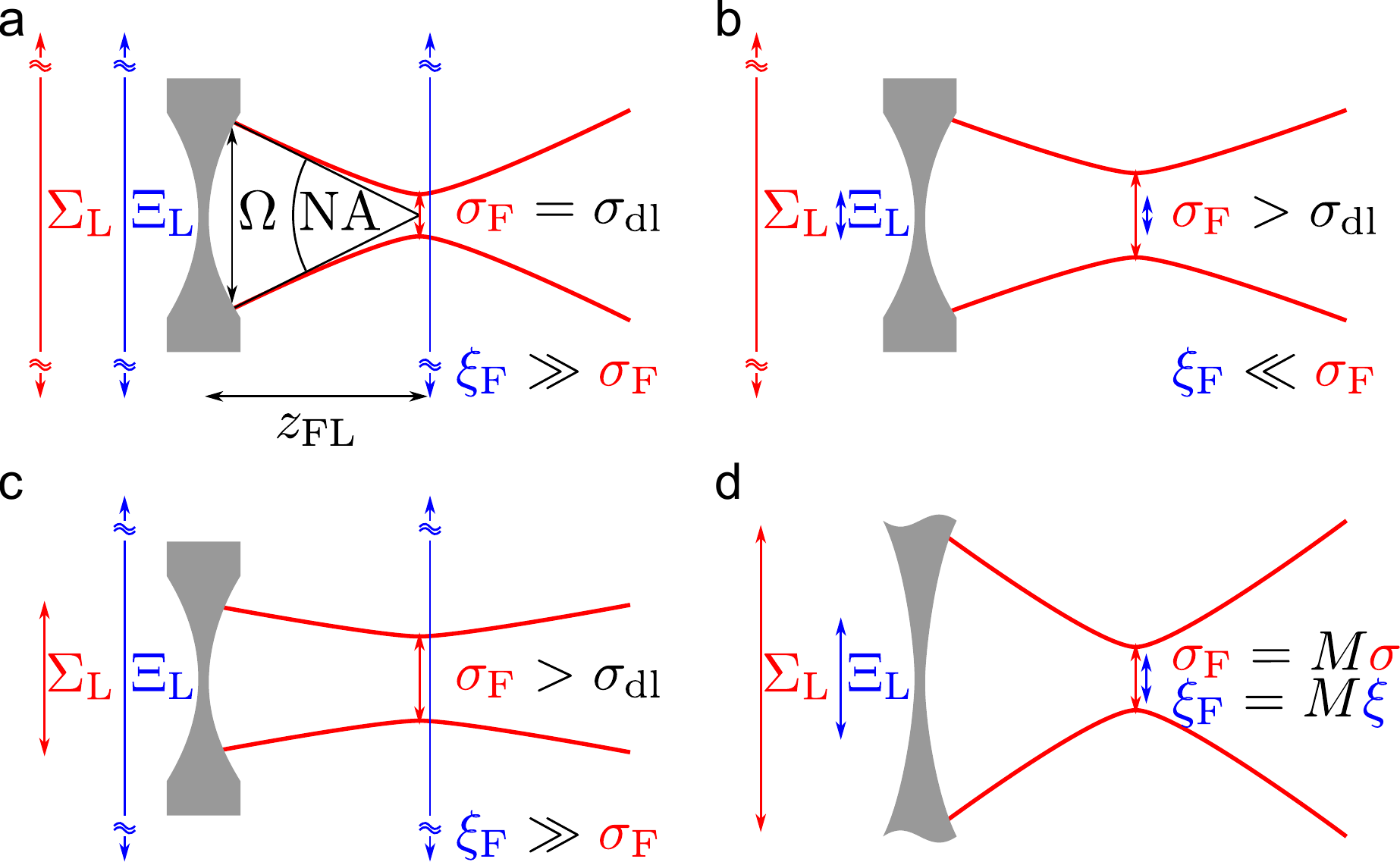} }
\caption{
(a) Focus is diffraction limited $\sigma_\textrm{F}=\sigma_\textrm{dl}$ if the beam size $\Sigma_\textrm{L}$ and transverse coherence length $\Xi_\textrm{L}$ are larger than the lens aperture $\Omega$.
(b,c) Focus size is larger than the diffraction limit if the coherence length (b) or beam size (c) incident on the lens is smaller than the aperture.
(d) For a very large aperture the focus is a demagnified image of the source.
}
\label{fig:SketchFocusCoherence}
\end{figure}

Using the concept of the diffraction limit (see equation \eqref{eq:DiffLim}) important cases for focusing of partially coherent radiation can be identified.
It can be immediately seen from equations (\ref{eq:SigmaBehind}, \ref{eq:DiffLim}) that the diffraction limit is the smallest possible focus size achievable with the lens, since $\tilde\Sigma_\textrm{L}\le\Omega$ and $\zeta_\textrm{F}\le1$ by definition.
It is also clear from equations (\ref{eq:SigmaBehind}, \ref{eq:zeta_F}, \ref{eq:DiffLim}) that the focus is diffraction limited only if both the beam size and transverse coherence length of the beam incident on the lens are much larger than the lens aperture (see Figure \ref{fig:SketchFocusCoherence} (a)).
The focus size increases if either the beam size or coherence length of the beam incident on the lens is smaller than the aperture of the lens (see Figure \ref{fig:SketchFocusCoherence} (b,c)).
However, there is an important difference between these two cases.
In the first example a highly coherent beam is obtained in the focus and the blurring of the focus size is due to diffraction effects of a finite incoming beam (see Figure \ref{fig:SketchFocusCoherence} (b)).
In the second case the beam in the focus is rather incoherent and the larger focus is a consequence of a small degree of coherence in the focus (see Figure \ref{fig:SketchFocusCoherence} (c)).

We can also express the focus size in terms of the beam parameters incident on the lens, which may be important for practical purposes.
Substituting (\ref{eq:SigmaBehind},~\ref{eq:zeta_F}) in equation \eqref{eq:DiffLim} we can obtain
\begin{equation}
  \sigma_\textrm{F}
  =\sigma_\textrm{dl}\cdot\left[ 1+\left( \frac{\Omega}{\Sigma_\textrm{L}} \right)^2
  + 4 \left( \frac{\Omega}{\Xi_\textrm{L}} \right)^2 \right]^{1/2}.
  \label{eq:sigma_fSFL}
\end{equation}
The coherence length in the focus can be calculated using the ratio $\xi_\textrm{F}/\sigma_\textrm{F}=\Xi_\textrm{L}/\tilde\Sigma_\textrm{L}$ (see equations (\ref{eq:sigma_f},~\ref{eq:xi_f},~\ref{eq:SigmaBehind})) and $\sigma_\textrm{F}$ from equation \eqref{eq:sigma_fSFL}
\begin{equation}
  \xi_\textrm{F}=\sigma_\textrm{F}\cdot\sqrt{1 +\frac{\Sigma_\textrm{L}^2}{\Omega^2}} \cdot\left(\frac{\Xi_\textrm{L}}{\Sigma_\textrm{L}}\right).
  \label{eq:xi_fSFL}
\end{equation}

\begin{figure}
\centering
\scalebox{1}{\includegraphics{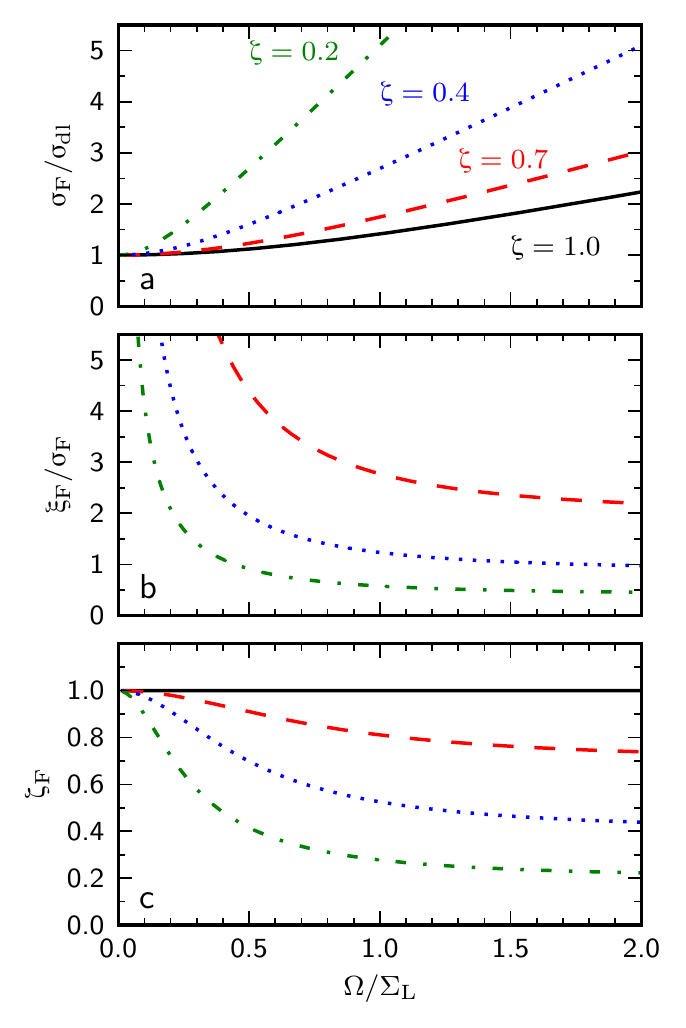} }
\caption{
The normalized focus size $\sigma_\textrm{F}/\sigma_\textrm{dl}$ (a), transverse coherence length $\xi_\textrm{F}/\sigma_\textrm{F}$ (b), and global degree of coherence $\zeta_\textrm{F}$ in the focus (c) as functions of the ratio $\Omega/\Sigma_\textrm{L}$.
Different values of the global degree of coherence at the source $\zeta=1.0$, 0.7, 0.4, and 0.2 are considered.
}
\label{fig:FocusSize}
\end{figure}

We demonstrate the obtained results in Figure \ref{fig:FocusSize}, where the focal size $\sigma_\textrm{F}$ \eqref{eq:sigma_f}, coherence length $\xi_\textrm{F}$ \eqref{eq:xi_f}, and degree of coherence $\zeta_\textrm{F}$ \eqref{eq:zeta_F} in the focus are calculated as a function of the ratio $\Omega/\Sigma_\textrm{L}$ for different values of the degree of coherence $\zeta$ of the incoming beam.
It is well seen from this figure that a diffraction limited focus size is obtained in the limit of a fully coherent beam.
With the reduced coherence of the incident beam the focal size is increased rapidly.
At the same time for the smaller apertures diffraction limit can be reached for beams of any degree of coherence, however, on the expense of limited photon flux.
Even for a highly coherent beam, the focus is larger than the diffraction limit if the beam size is smaller than the lens aperture.
As it is well seen from Figure \ref{fig:FocusSize} (b) the ratio of the transverse coherence length to the focus size $\xi_\textrm{F}/\sigma_\textrm{F}$ increases rapidly and approaches infinity for smaller apertures.
As the focal size approaches the diffraction limit the degree of coherence approaches the fully coherent value of $\zeta_\textrm{F}=1$ at small apertures (see Figure \ref{fig:FocusSize} (c)).

A summary of equations obtained in this section to determine the beam properties in the focus is given in Table \ref{tab:lens2}.

\bigskip
\begin{table}
\centering
\caption{Coherence properties in the focus of a strongly focusing lens $\tilde R_\textrm{L}\ll Z_\textrm{L}$ with an arbitrary lens aperture.
}
  \begin{tabular}{l|l}
    \textrm{Focus size}\qquad&$ \sigma_\textrm{F} =\sigma_\textrm{dl}\left[ 1+\left( {\Omega}/{\Sigma_\textrm{L}} \right)^2
      +4\left( {\Omega}/{\Xi_\textrm{L}} \right)^2 \right]^{1/2}$\\
      \textrm{Diffraction limit}\qquad&$\sigma_\textrm{dl}=z_\textrm{FL}/(2k\Omega)$\\
    \textrm{Transverse coherence length}\qquad&$\xi_\textrm{F}=\sigma_\textrm{F} \cdot\sqrt{1+\Sigma_\textrm{L}^2/\Omega^2}\cdot( \Xi_\textrm{L}/\Sigma_\textrm{L})$\\
    \textrm{Focus Position}\qquad&$z_\textrm{FL}=R_\textrm{L}f/(R_\textrm{L}-f)$\\
    \textrm{Depth of focus}\qquad&$\Delta f = 4k\sigma_\textrm{F}^2\zeta_\textrm{F}$
  \end{tabular}
  \label{tab:lens2}
\end{table}
\bigskip

\section{Focusing element with a large aperture}


If the lens aperture $\Omega$ is significantly larger than the beam size of the incident radiation $\Sigma_\textrm{L}$, the lens only modifies the radius of curvature.
Then the beam size and coherence length in the focus can be expressed through the same parameters at the source through simple relations \cite{TF1986}%
\footnote{These expressions hold even without the strong lens approximation used earlier in the paper and can be obtained from equations (\ref{eq:sigma_f},~\ref{eq:xi_f}) by a straightforward calculation \cite{SingerThesis}.}
\begin{equation}
	\sigma_\textrm{F}=M\sigma,\qquad \xi_\textrm{F}=M\xi,
\label{eq:focusopt}
\end{equation}
where
\begin{equation}
  M=\left|\dfrac{f}{z_{\textrm{L}0}-f}\right|\left( 1+\dfrac{z_\textrm{eff}^2}{(z_{\textrm{L}0}-f)^2} \right)^{-1/2}
\end{equation}
is the magnification factor (see Figure \ref{fig:SketchFocusCoherence} (d)).
The ratio between the transverse coherence length and the beam size is constant everywhere along the optical axis and is determined by the source parameters $\xi/\sigma$.
The same holds for the degree of transverse coherence $\zeta$.
As an important result we note that in the frame of the GSM the focus generated by a CRL with a sufficiently large aperture is just a scaled image of the source.

In the limit of geometrical optics, when diffraction effects can be neglected and the degree of coherence approaches zero $(\zeta\to0)$ \cite{BW1999,G2005} the effective distance vanishes $z_\textrm{eff}\to 0$ and magnification factor simplifies to
\begin{equation}
  M=\left|\dfrac{f}{z_{\textrm{L}0}-f}\right|.
\label{eq:mag}
\end{equation}
The same limit is approached if the distance $z_\textrm{L0}-f \gg z_\textrm{eff}$, which is typical for synchrotron sources.
A summary of the equations applicable for the case of large apertures is presented in Table \ref{tab:lens1}
\bigskip
\begin{table}
  \centering
  \caption{Coherence properties in the focus of a lens with an aperture much larger than the beam size $\Omega\gg\Sigma_\textrm{L}$}
  \begin{tabular}{c|c}
    \textrm{Focus size}\qquad&$\sigma_\textrm{F}=M\sigma$\\
    \textrm{Transverse coherence length}\qquad&$\xi_\textrm{F}=M\xi$\\
    \textrm{Focus position}\qquad&$z_\textrm{FL}=f+M^2(z_{\textrm{F}0}-f)$\\
    \textrm{Depth of focus}\qquad&$\Delta f=4 k M^2\sigma^2\zeta$\\
    \textrm{Magnification}\qquad&$M=\left|\dfrac{f}{f-z_{\textrm{L}0}}\right|\left( 1+\dfrac{(2k\sigma^2\zeta)^2}{(f-z_{\textrm{L}0})^2} \right)^{-1/2}$
  \end{tabular}
  \label{tab:lens1}
\end{table}

\section{Focusing of X-ray beams at 3rd generation synchrotron sources}

We have applied the general approach developed in the previous sections to simulate the coherence properties of the focused X-ray beams at the beamline P10 at PETRA III.
This beamline is dedicated for coherence applications such as CXDI and XPCS and understanding of the coherence properties in the focus is vital for the success of these experiments.

As an example we analyzed an optical system installed at this beamline, which consists of three Berylium CRLs with radii of 200 $\mu$m, 50 $\mu$m, and 50 $\mu$m and is positioned at a distance of 85 m downstream of the source \cite{ZozulyaOptExp2012}.
We considered this set of lenses as a thin lens and applied equations (\ref{eq:f},\ref{eq:Omegatot}) to determine the focal length $f=2.13$ m and the effective aperture of the lens due to absorption $\Omega_0=242~\mu$m.
The geometrical size of the 50 $\mu$m lenses is $450~\mu$m%
\footnote{We should note that the geometrical lens size limits the aperture $\Omega_\textrm{A}$.
According to our estimates it corresponds to an rms width of about 100 $\mu$m (see Appendix \ref{chap:A1}).}.
\bigskip
\begin{table}
\centering
\caption{Beam parameters of the PETRA III source (low-$\beta$) for a photon energy of 8 keV.
The coherence properties at the source and at a distance of 85 m downstream of the source are presented \cite{VartanyantsNJP2010}.
}
\begin{tabular}{c|c|c}
&Horizontal&Vertical\\
Beam size at the source, $\mu$m 			&36.2 	&6.3\\
Transverse coherence length at the source, $\mu$m 	&0.9 	&7.7\\
Beam size at 85 m, $\mu$m				&2370 	&320\\
Transverse coherence at 85 m, $\mu$m			&58	&390\\
Degree of coherence $\zeta$				&0.01	&0.52\\
\end{tabular}
\label{tab:beamparameters}
\end{table}
\bigskip

We analyzed the coherence properties of such a lens as a function of the aperture size $\Omega_\textrm{A}$.
To determine the beam properties in the region near the focal plane we have used the general expression (\ref{eq:GSM_pfs}).
The parameters of the source were considered for a photon energy of 8 keV and low-$\beta$ operation of the synchrotron source (see Table \ref{tab:beamparameters}).
It is immediately seen that the radiation in the horizontal and vertical directions can be considered as incoherent and coherent, respectively (see also Figure \ref{fig:FocusSize}).
Equation \eqref{eq:Omegatot} was used to calculate the total aperture size $\Omega$ of the focusing element including the beam defining aperture.
Aperture sizes $\Omega_\textrm{A}$($\Omega$) of 25 (25) $\mu$m and 100 (93) $\mu$m were considered in the horizontal and 50 (49) $\mu$m and 150 (128) $\mu$m in vertical direction (see Table \ref{tab:lensP10}).

\begin{figure}
\centering
\scalebox{.165}{\includegraphics{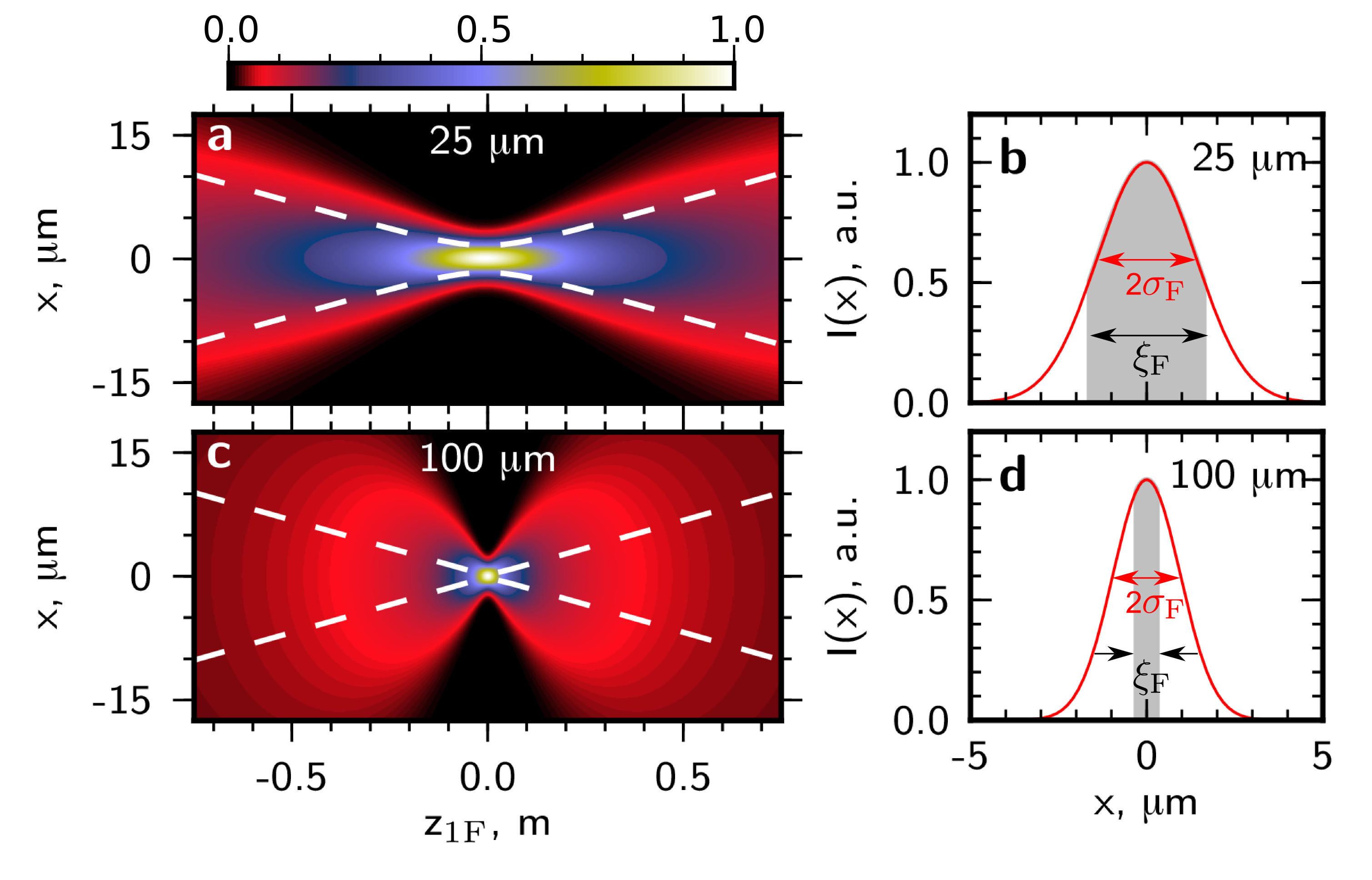} }
\caption{The intensity profile in the vicinity of the focus as a function of the propagation distance $z_\textrm{1F}$ for the aperture sizes $\Omega_\textrm{A}$ of 25 $\mu$m (a), 100 $\mu$m (c) in horizontal direction.
The white dashed line indicates the coherent part of the beam with a width given by the transverse coherence length.
(b,d) Line scans of the intensity profile $I(x)$ from (c,d) in the focal plane at $z_\textrm{1F}=0$.
The shaded region shows the coherent part of the beam with the width corresponding to the transverse coherence length $\xi_\textrm{F}$ in the focal plane.
}
\label{fig:hor_Gauss}
\end{figure}
\begin{figure}
\centering
\scalebox{0.165}{\includegraphics{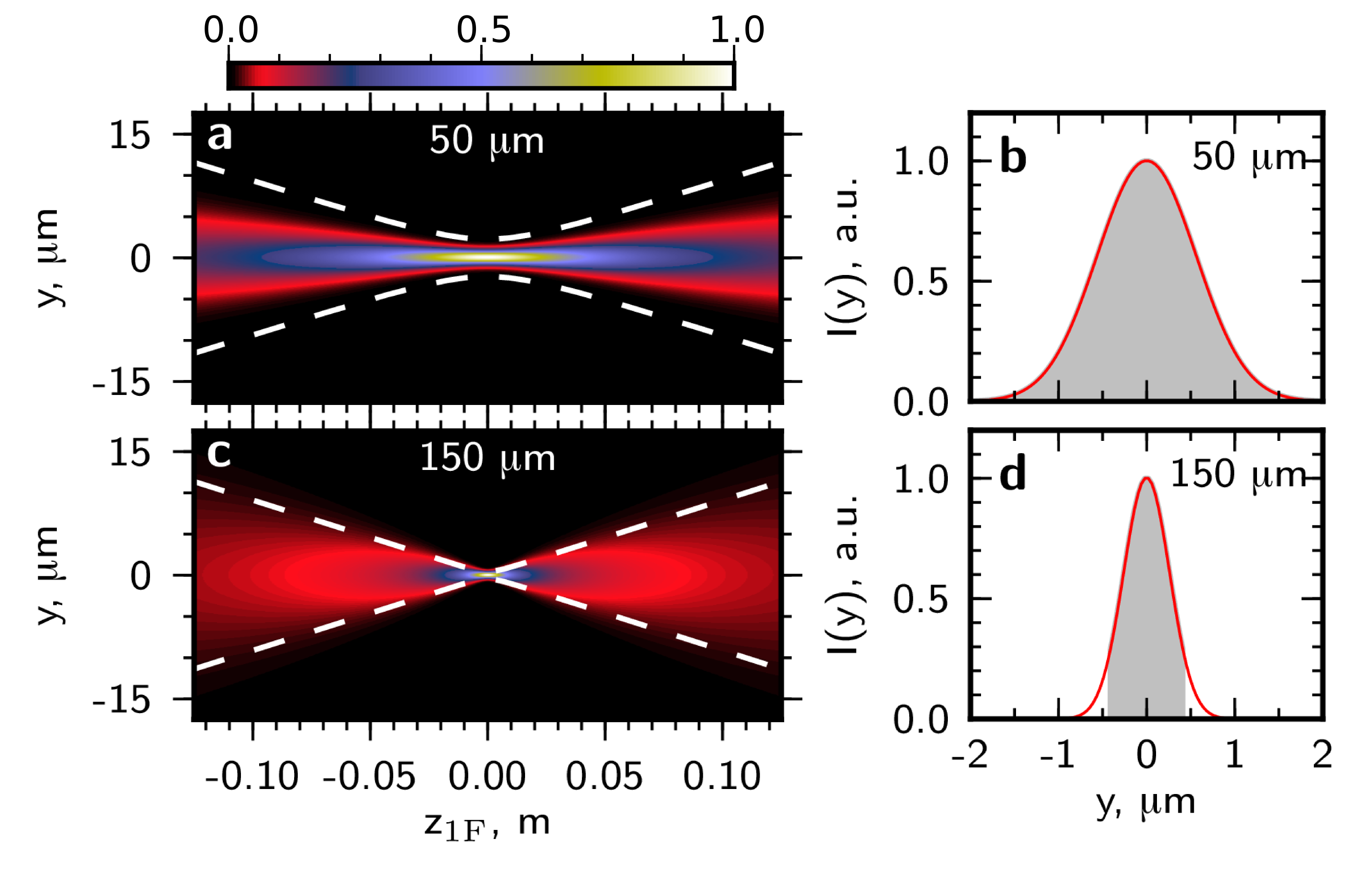}}
\caption{The same as in Figure \ref{fig:hor_Gauss} in the vertical direction and for the aperture sizes $\Omega_\textrm{A}$ of 50 $\mu$m (a,b), 150 $\mu$m (c,d).
}
\label{fig:ver_Gauss}
\end{figure}

In Figures \ref{fig:hor_Gauss} and \ref{fig:ver_Gauss} the intensity profile and transverse coherence properties at different distances from the lens around the focal position in the horizontal and vertical directions are presented.
In the horizontal direction for an aperture size of 25 $\mu$m the coherence length is about twice times larger than the beam size and the beam is highly coherent (see Figure \ref{fig:hor_Gauss} (a,b)).
The depth of focus is about ten centimeters.
For a significantly larger aperture size of 100 $\mu$m the focus size and the depth of focus are smaller and the beam coherence is poor (see Figure \ref{fig:hor_Gauss} (c,d)).
In the vertical direction the beam size and the depth of focus are significantly smaller than for the horizontal direction (note different scales in Figures \ref{fig:hor_Gauss} and \ref{fig:ver_Gauss}).
Due to a higher degree of coherence at the source in the vertical direction highly coherent radiation in the focus can be achieved with larger apertures.
For the aperture size of 50 $\mu$m the coherence length is significantly larger than the beam size and the beam is fully coherent (see Figure \ref{fig:ver_Gauss} (a,b)).
Even for a comparably large aperture of 150 $\mu$m the coherence length substantially exceeds the beam size (see Figure \ref{fig:ver_Gauss} (c,d)).

\begin{figure}
\centering
\scalebox{1}{\includegraphics{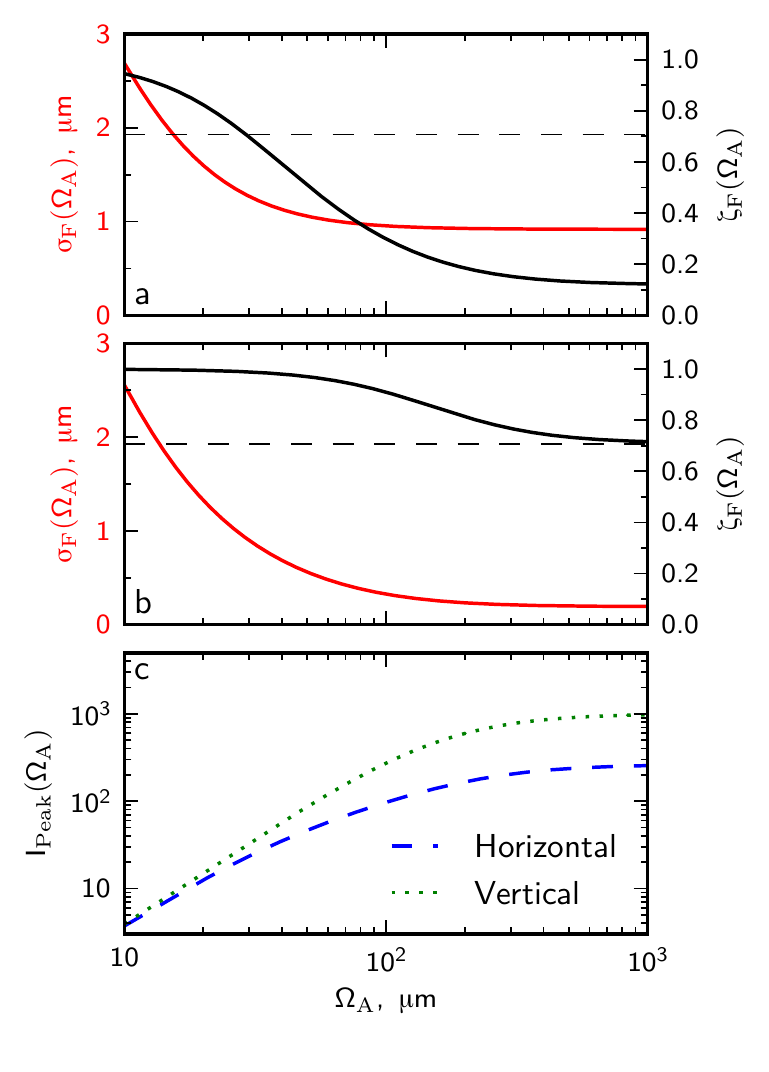} }
\caption{
The global degree of coherence (black line) and the focal size (red line) as functions of the beam defining aperture size $\Omega_\textrm{A}$ in the horizontal (a) and vertical (b) directions.
The dashed line indicates a highly coherent beam with the coherence length in the focus being twice larger than the beam size $\xi_\textrm{F}=2\sigma_\textrm{F}$.
(c) The increase in the flux density $I_\textrm{Peak}$ as a function of the beam defining aperture size in the horizontal (blue dashed line) and vertical (green dotted line) directions.
}
\label{fig:focus_coherence}
\end{figure}

For coherence based applications the most important properties of an X-ray lens are the small focus size, increase of the degree of coherence in the focus, and increase in the peak intensity $I_\textrm{Peak}=\textrm{max}\{I(x,y)\}$.
These quantities as functions of the opening aperture size $\Omega_\textrm{A}$ are presented in Figure \ref{fig:focus_coherence} for the horizontal and vertical directions%
\footnote{For this set of lenses aperture sizes of more than 100 $\mu$m are larger than the geometrical size of the lens and are shown to illustrate the asymptotic behaviour.}.
As well seen in the Figure \ref{fig:focus_coherence} (a,b) at the largest apertures both the focus size and degree of coherence have constant values and the smallest focus size is obtained.
The focus size $\sigma_\textrm{F}$ is increased at smaller aperture values $\Omega_\textrm{A}$ due to diffraction as described in section 4.
At the same time the degree of coherence $\zeta_\textrm{F}$ reaches its maximum value close to one.
In the horizontal direction it increases from a value of 10\% with large beam defining aperture $\Omega_\textrm{A}$ to 71\% (horizontal dashed line in Figure \ref{fig:focus_coherence} (a)) for an aperture size of about 30 $\mu$m.
This can be considered as a highly coherent beam with the coherence length being twice the size of the beam.
In the vertical direction the degree of coherence is higher than 71\% for all aperture sizes of the optical system considered here.
The peak intensity in the focus theoretically can be increased by more than two orders of magnitude in both directions for large apertures (see Figure \ref{fig:focus_coherence} (c)).
At the same time for the small aperture sizes the amount of the total transmitted flux is reduced.
In this focusing geometry using an aperture sizes of 30 $\mu$m (H) and 100 $\mu$m (V) a highly coherent beam with the focus size of 1.2 $\mu$m (H) and 0.3 $\mu$m (V) is expected.
In this case 0.2\% of the total flux is transmitted through the lens and the flux density is increased by three orders of magnitude.

\bigskip
\begin{table}
\centering
\caption{Coherence properties in the focus of the beamline P10 at PETRA III calculated for different apertures $\Omega_\textrm{A}$ in front of the lens.
}
\begin{tabular}{lcccc}
						&\multicolumn{2}{c}{Horizontal}&\multicolumn{2}{c}{Vertical}\\
Aperture size $\Omega_\textrm{A}$, $\mu$m		&25 		&100 		&50 		&150\\
Total aperture size $\Omega$, $\mu$m			&25 		&93 		&49 		&128\\
Focus size $\sigma_\textrm{F}$, $\mu$m			&1.4		&1.0		&0.6		&0.3\\
Transverse coherence length $\xi_\textrm{F}$, $\mu$m	&3.3		&0.6		&4.5		&0.9\\
Global degree of coherence $\zeta_\textrm{F}$ 		&0.76 		&0.30 		&0.97		&0.85\\
Depth of focus $\Delta f$, mm			 	&120		&22 		&25		&3.6\\
\end{tabular}
  \label{tab:lensP10}
\end{table}
\bigskip


We have compared the results of our approach with the measurements of the beam size performed at the coherence beamline P10 \cite{ZozulyaOptExp2012}.
A transfocator with seven 50 $\mu$m Berylium CRLs was used at an energy of 13.2 keV.
The beam defining slits were set to 100 $\mu$m in both directions and a focus size (FWHM) of 2.9 $\mu$m (H) x 2.9 $\mu$m (V) was observed.
Applying our approach for the same lens parameters ($\Omega_\textrm{A}=22$) and the estimated source properties at a photon energy of 13.2 keV \cite{VartanyantsNJP2010} yields a theoretical focus size (FWHM) of 2.4 $\mu$m (H) x 1.5 $\mu$m(V).
Our simulations reproduce well experimental focus size in the horizontal direction, however, they are about twice smaller than the measured values in the vertical direction.
This can be attributed to the fact that all optical components at P10 deflect the beam in the vertical direction and we expect the deviation of the experimental and theoretical values to be larger in this direction.

\section{Conclusions}

We have presented an analytic approach to propagate partially coherent X-ray beams through focusing elements, which is based on the results of statistical optics and can be applied to X-Ray beams at third generation synchrotron sources.
As an example parabolic compound refractive lenses were analyzed in detail.
The same formalism can be also applied to Fresnel zone plates and other focusing optics, which can be treated within the thin lens approximation.
We have obtained simple equations for the case of a strongly focusing lens.
Since the method is analytical it can be effectively used to estimate the beam parameters at the experimental station.
Important limiting cases, such as rather coherent and incoherent radiation have been considered, which represent synchrotron radiation in the vertical and horizontal direction, respectively.
As an example we have performed calculations for the coherence beamline P10 at PETRA III storage ring.
We anticipate that our approach can be also applied to estimate the performance of focused beams at highly coherent X-ray free-electron laser sources.

\section{Acknowledgements}

We acknowledge fruitful discussions with E. Weckert and careful reading of the manuscript by M. Sprung and H. Franz.
We also acknowledge discussions with A. Zozulya concerning the P10 focusing optics.
Part of this work was supported by BMBF Grant No. 5K10CHG ''Coherent Diffraction Imaging and Scattering of Ultrashort Coherent Pulses with Matter'' in the framework of the German-Russian collaboration ''Development and Use of Accelerator-Based Photon Sources'' and the Virtual Institute VH-VI-403 of the Helmholtz Association.

\appendix

\section{Edge effects of a beam defining slit on the coherence properties in the focus}

\label{chap:A1}
A potential problem of the application of the Gaussian beam defining aperture to the case of synchrotron radiation sources is the fact that at the most beamlines a slit or a pinhole is used, which has a non Gaussian transmission function.
To understand how significant the edges of such apertures can be, we performed numerical propagation of the CSD from the lens with an aperture in the form of a slit with the size $D$ (the case of the pinhole was considered in \cite{SV2011}).
Equations (\ref{eq:CSD_propagation}, \ref{eq:Fresnel_propagator}, \ref{eq:CSD_thin_lens}) were solved numerically using the following transmission function of the optical system
\begin{equation}
	T(u) = T_\textrm{S}(u)\cdot\exp\left(-\frac{u^2}{4\Omega_0^2}-i\frac{ku^2}{2f}\right),
\end{equation}
where $T_\textrm{S}(u)=1$ if $|u|<D/2$ and $0$ elsewhere.
The lens aperture due to absorption $\Omega_0$ and the focal length of the lens $f$ were the same as in the main text.
The slit transmission function $T_\textrm{S}(u)$ was convolved by a Gaussian with a width of 20 $\mu$m, to smooth the hard edges.
These simulations were compared with our analytical approach.
The size of the Gaussian aperture $\Omega_\textrm{A}$ was related to the slit size $D$ by a comparison of the intensity profiles (FWHM) generated by a Gaussian and a rectangular transmission function.
The best match was found for the condition $D=4.55\cdot\Omega_\textrm{A}$.

\begin{figure}
\centering
\includegraphics{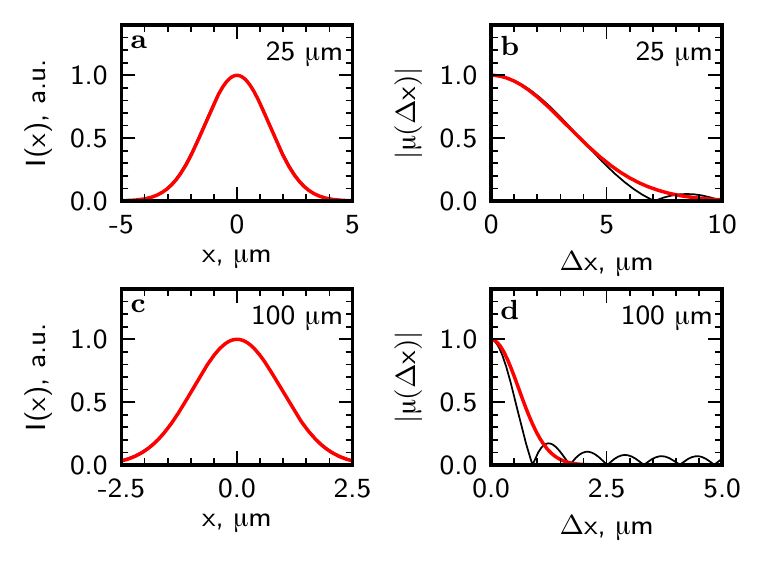}
\caption{The intensity profile $I(x)$ (a,c) and modulus of the spectral degree of coherence $|\mu(\Delta x)|$ (b,d) in the focal plane in the horizontal direction.
Calculations made for the aperture sizes of 25 $\mu$m (a) and 100 $\mu$m (b) with a Gaussian aperture (red lines) and a slit (black line) are presented.}
\label{fig:hor_focus}
\end{figure}

\begin{figure}
\centering
\includegraphics{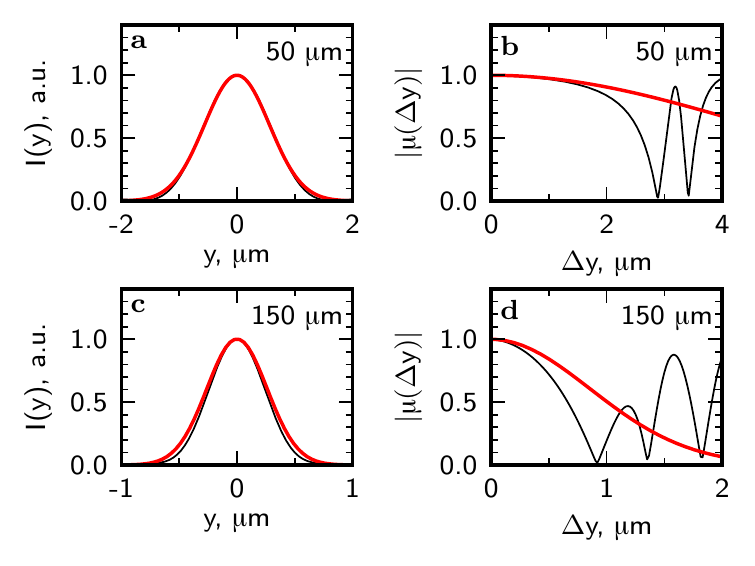}
\caption{The same as in Figure \ref{fig:hor_Gauss} in the vertical direction for the aperture sizes of 50 $\mu$m (a,c) and 150 $\mu$m (b,d).}
\label{fig:ver_focus}
\end{figure}

In Figures \ref{fig:hor_focus} and \ref{fig:ver_focus} the intensity profile and the spectral degree of coherence in the focus are shown for the same aperture sizes as in Figures \ref{fig:hor_Gauss} and \ref{fig:ver_Gauss}.
For comparison the corresponding intensity profiles and SDC obtained in numerical simulations are shown in the same Figures.

It is well seen, that the beam profile determined through the analytical model presented in this work coincides well with the results of numerical simulations.
Apart from oscillations in low intensity regions, which appear due to diffraction on hard edges of the slit, the Gaussian model describes well the coherence properties of the focused beam in the horizontal direction.
In the vertical direction the edge effects are substantial due to a high coherence, and the GSM slightly overestimates the coherence length of the beam.

\section{Derivation of the focal distance, focus size, and transverse coherence length}

\label{chap:A2}
To determine the radiation properties in the focus we use the optics reciprocity theorem.
We consider the focus as a GSM source and the radiation behind the lens as back propagated from the focus.
We have to determine the focus size $\sigma_\textrm{F}$, transverse coherence length $\xi_\textrm{F}$, and distance from the lens to the focus $z_\textrm{FL}$ from the beam size $\tilde \Sigma_\textrm{L}$, coherence length $\Xi_\textrm{L}$, and radius of curvature $\tilde R_\textrm{L}$ immediately behind the lens.
These parameters are connected through the expressions for the expansion coefficient $\Delta_\textrm{L}$ (\ref{eq:GSM_pfs}, \ref{eq:SigXi})
\begin{eqnarray}
	\tilde\Sigma_\textrm{L}=\sigma_\textrm{F}\Delta_\textrm{L},\quad
	\Xi_\textrm{L}=\xi_\textrm{F}\Delta_\textrm{L},\quad
\Delta_\textrm{L}=\left[ 1+ \left(\frac{z_\textrm{FL}}{z^\textrm{F}_\textrm{eff}} \right)^2\right]^{1/2}
\end{eqnarray}
and the radius of curvature
\begin{equation}
	\tilde R_\textrm{L}= -z_\textrm{FL}\left[1+\left(\frac{z^\textrm{F}_\textrm{eff}}{z_\textrm{FL}} \right)^2  \right].
\end{equation}
To solve this set of equations we rewrite the expansion coefficient $\Delta_\textrm{L}$ in terms of the radius of curvature
\begin{equation}
	\Delta_\textrm{L}^2 = -\frac{\tilde R_\textrm{L} z_\textrm{FL}}{{z_\textrm{eff}^\textrm{F}}^2}.
\label{eq:A0}
\end{equation}
Introducing a new variable $Z_1 = \Delta_\textrm{L}^2z_\textrm{eff}^\textrm{F}=2k\tilde\Sigma_\textrm{L}^2\zeta_\textrm{F}$, which can be calculated from the parameters of the incident radiation and the lens aperture using equations (\ref{eq:SigmaBehind}, \ref{eq:zeta_F}, \ref{eq:A0}) we rewrite
\begin{equation}
\Delta_\textrm{L}^2 = - \frac{Z_1^2}{\tilde R_\textrm{L}z_\textrm{FL}}
\label{eq:A1}
\end{equation}
and
\begin{equation}
\Delta_\textrm{L}^2 = 1 + \Delta_\textrm{L}^4\left(\frac{z_\textrm{FL}}{Z_1}\right)^2.
\label{eq:A2}
\end{equation}
Now substituting \eqref{eq:A1} in \eqref{eq:A2} we find the distance from the lens to the focus
\begin{equation}
z_\textrm{FL}=-\frac{\tilde R_\textrm{L}}{1+\left({\tilde R_\textrm{L}}/Z_1\right)^2}
\label{eq:A3}
\end{equation}
Substituting equation \eqref{eq:A3} in \eqref{eq:A1} we find the expansion coefficient
\begin{equation}
\Delta_\textrm{L}^2 = 1 + \left(\frac{Z_1}{\tilde R_\textrm{L}}\right)^2.
\end{equation}

\section{Focus size as a function of the source size}
\label{chap:A3}
Here we show that in the strong focusing approximation the focus size can be given as a convolution of the diffraction limit $\sigma_\textrm{dl}$ and demagnified source size $M\sigma$, where $M$ is the demagnification factor.
Substituting the definition of the diffraction limit $\sigma_\textrm{dl}=z_\textrm{FL}/(2k\Omega)$ in equation \eqref{eq:sigma_fSFL} we find
\begin{equation}
\sigma_\textrm{F}^2
  =\sigma_\textrm{dl}^2
	+ \left(\frac{z_\textrm{FL}}{2k\Sigma_\textrm{L}}\right)^2
	+ \left(\frac{z_\textrm{FL}}{k\Xi_\textrm{L}}\right)^2.
\end{equation}
Now using equations (\ref{eq:SigXi}, \ref{eq:Delta}, \ref{eq:zeff}, \ref{eq:zeta_GSM}) in the far field approximation $z_\textrm{L0}\gg z_\textrm{eff}$ we find
\begin{equation}
	\left(\frac{1}{2\Sigma_\textrm{L}}\right)^2+\left(\frac1{\Xi}\right)^2
	= \left(\frac{k\sigma}{z_\textrm{L0}}\right)^2.
\end{equation}
The focus size can then be given by
\begin{equation}
\sigma_\textrm{F}^2=\sigma_\textrm{dl}^2+M^2\sigma^2,
\label{eq:demsource}
\end{equation}
with the demagnification factor expressed as $M=z_\textrm{FL}/z_\textrm{L0}$.
In the strong focusing approximation $Z_\textrm{L}\gg \tilde R_\textrm{L}$ using equation (\ref{eq:RBehind}, \ref{eq:FocalPosition}) the demagnification factor can be given as
\begin{equation}
M = \left\vert\frac{f}{f-z_\textrm{L0}}\right\vert.
\end{equation}

It is interesting to note, that when the beam is incoherent $\Xi_\textrm{L}\ll\Sigma_\textrm{L}$ and the aperture is larger than the transverse coherence length $\Omega\gg\Xi_\textrm{L}$, we find from equation \eqref{eq:sigma_fSFL} $\sigma_\textrm{F}=z_\textrm{FL}/(k\Xi_L)$ (see Figures \ref{fig:FocusSize} and \ref{fig:SketchFocusCoherence} (b)).
In these conditions the focus size is determined only by the transverse coherence length of the beam incident on the lens.
Rewriting the condition for the focus size to $\Xi_\textrm{L}=z_\textrm{FL}/(k\sigma_\textrm{F})$ it can readily be seen that this case is very similar to the van Cittert-Zernike theorem \cite{MW1995}.
The focus can be considered as a planar incoherent GSM source and the transverse coherence length at a distance $z_\textrm{FL}$ from this source is given by $\Xi_\textrm{L}$.
In fact, the coherence length is demagnified in the focus by the lens.

\end{document}